\documentclass[preprint,12pt]{elsarticle}

\usepackage{amssymb}
\usepackage[utf8]{inputenc} % allow utf-8 input
\usepackage[T1]{fontenc}    % use 8-bit T1 fonts
\usepackage{hyperref}       % hyperlinks
\usepackage{url}            % simple URL typesetting
\usepackage{booktabs}       % professional-quality tables
\usepackage{amsfonts}       % blackboard math symbols
\usepackage{nicefrac}       % compact symbols for 1/2, etc.
\usepackage{microtype}      % microtypography
\usepackage{lipsum}
\usepackage{graphicx}       % include graphics
\journal{Applied Surface Science}
\begin{document}

\begin{frontmatter}

\title{Complementary electrochemical ICP-MS flow cell and in-situ AFM study of the anodic desorption of molecular adhesion promotors}

\author[tuaip]{Dominik Dworschak}
\address[tuaip]{Institute of Applied Physics, Vienna University of Technology, Vienna, Austria}

\author[tuaip]{Carina Brunnhofer}
\author[tuaip]{Markus Valtiner}
\ead{markus.valtiner@tuwien.ac.at}

% \begin{document}
% \maketitle

\begin{abstract} % 200 words allowed, currently 160
Molecular adhesion promoters are a central component of modern coating systems for the corrosion protection of structural materials. 
They are interface active and form ultrathin corrosion inhibiting and adhesion-promoting layers. Here we utilize thiol-based self-assembled monolayers (SAMs) as model system for demonstrating a comprehensive combinatorial approach to understand molecular level corrosion protection mechanisms under anodic polarization.  
Specifically, we compare hydrophilic 11-Mercapto-1-undecanol and hydrophobic 1-Undecanethiol SAMs and their gold-dissolution inhibiting properties. We can show that the intermolecular forces (hydrophobic vs hydrophilic effects) control how SAM layers perform under oxidative conditions. Specifically, using \textit{in situ} electrochemical AFM and a scanning-flow cell coupled to an ICP-MS a complementary view on both corrosion resistance, as well as on changes in surface morphology/adhesion of the SAM is possible. Protection from oxidative dissolution is higher with hydrophobic SAMs, which detach under micelle formation, while the hydrophilic SAM exhibits lower protective effects on gold dissolution rates, although it stays intact as highly mobile layer under anodic polarization. 
The developed multi-technique approach will prove useful for studying the interfacial activity and corrosion suppression mechanism of inhibiting molecules on other metals and alloys.

\end{abstract}

\begin{graphicalabstract}
\includegraphics[width = \textwidth]{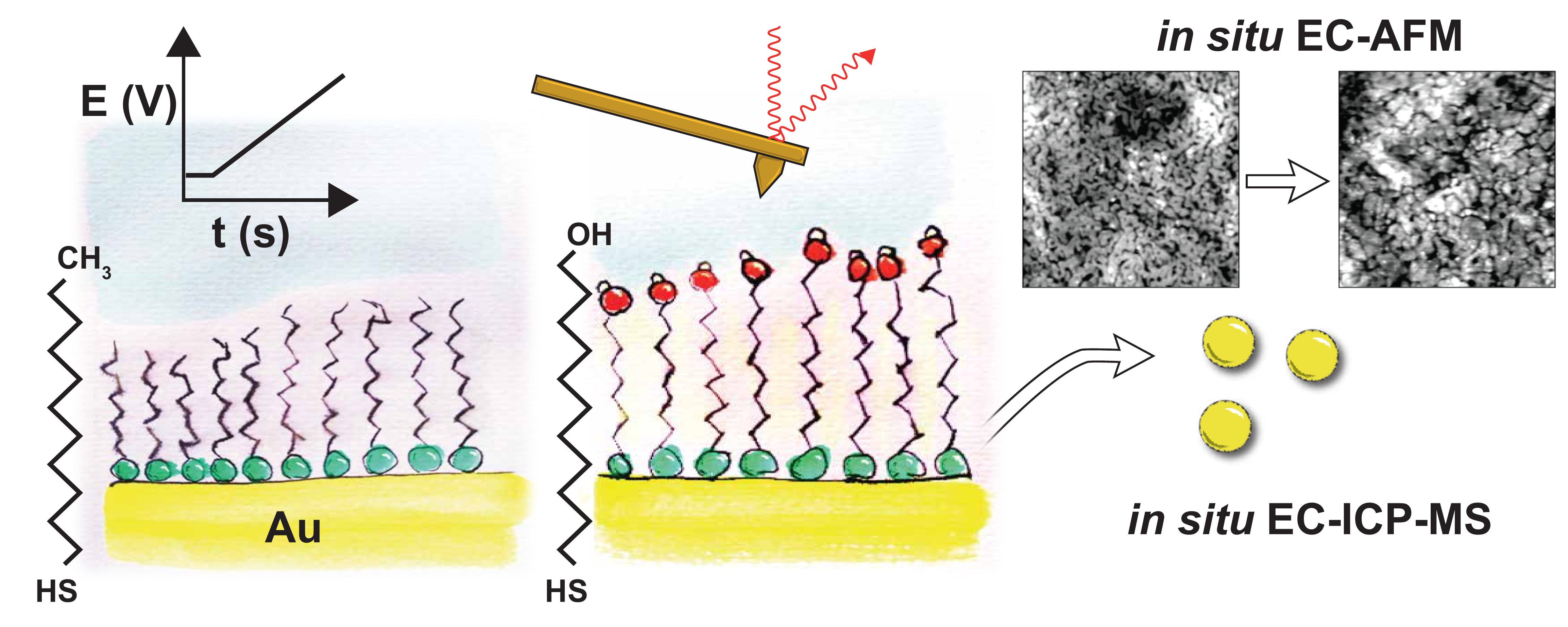}
\end{graphicalabstract}

\begin{highlights}
\item Anodic corrosion protection of SAMs on Au with varying hydrophobicity are compared.
\item Intermolecular forces control the behaviour of SAMs during anodic polarization
\item Hydrophobic SAM enhances corrosion resistance via micelle formation at interface.
\item Hydrophilic SAM remains weakly surface bound as a highly mobile layer. 

\end{highlights}

% keywords
\end{frontmatter}

\section{Introduction}
Organic coatings are widely used as corrosion protecting layers on metals\cite{Azzaroni.2001}.
Coatings protect from corrosion by a physical barrier effect for water between the metal and its surrounding, and are therefore up to a few micrometer in thickness \cite{Marcus.2012}.
Still, the most crucial part of the coating is at the boundary of the organic top coating with the metal(oxide) underneath \cite{Possart.2019b}.
Additionally, molecules such as silanes, or phosphonates \cite{Thissen.2010} which can form self assembling molecular thin films/monolayers (SAM) at an interface, are typical reactive additives for coating formulations. 
Specifically, SAMs have been proven to be suitable as molecular adhesion promotors for long-term corrosion protection \cite{Grundmeier.1998}. Further, the covalent character of the head-group to metal(oxide) bond stabilizes as a linking layer between the metal(oxide) and the further covalent attachment of polymer layers on metallic substrates.

In particular, the system of gold and thiol-based self-assembled monolayers (SAMs) is well studied as model system due to its ease of preparation, and stability and quality of the layers \cite{Vericat.2010,Srisombat.2011} .
For thiol-based SAMs the bond strength \cite{Grandbois.1999} between the sulfur of the thiol and the gold determines the stability of the formed monolayer.
Potential dependent SAM formation has been measured \textit{ex situ} with Scanning tunneling microscopy (STM) \cite{Grundmeier.1998}, but not yet under operating conditions at oxidative/anodic potentials. 
Further, weakening of the bond for removal of the formed SAM by cathodic desorption (stripping) is a well established process \cite{Liu.2008,Kakiuchi.2001,Corthey.2009,Wano.2001,Vela.2000}. 

The oxidative desorption on the contrary is less studied \cite{Ovchinnikova.2015}.
Though electrochemical anodic detachment already finds application in formation of mixed SAM layers by partial electrochemically induced desorption of a SAM in a thiol solution \cite{Chen.2010}.
Further, polarization of the surface can enhance the adsorption kinetics of molecular thin films \cite{Liu.2008}. How molecular adhesion promotors act, and to which degree they suppress metal dissolution, when the underlying substrate is electrochemically polarized into anodic polarization is still unclear.

In recent years, elementally resolved electrochemical techniques such as combinations of flow cells with downstream inductively coupled plasma mass spectrometry (ICP-MS) or optical emission spectrometry (ICP-OES) analysis, proved to be useful for understanding molecular dissolution mechanisms at electroactive interfaces \cite{Ogle.2000,Ogle.2019,Dworschak.2020,Cherevko.2013}.
These techniques have been applied to study the dissolution of Zn coatings on steel \cite{Vu.2012,Jiang.2012}, the dissolution of electrocatalysts under operating conditions \cite{Cherevko.2013,Klemm.2012}, or to the study of the potential dependent stability of photo-electrocatalytic materials \cite{Dworschak.2020,Knoppel.2018}.
Here, we extend this technique to understand corrosion inhibiting effects of molecular adhesion promotors under anodic polarization, as a model system for more complex alloys and inhibitors. 

Specifically, we report results on the \textit{in situ} anodic polarization of SAM covered gold in sodium perchlorate solution. We compare Linear Sweep Voltammetry in hand with \textit{in situ} electrochemical topography scanning with an atomic force microscope (EC-AFM) and elementally resolved ICP-MS dissolution currents of hydrophobic and hydrophilic SAMs. We demonstrated a significant corrosion-inhibiting effect of SAMs at anodic potentials, and show how interfacial hydrophobic forces can result in a micelle formation at the interface. 
This comprehensive approach combines complementary interface sensitive techniques for providing a detailed insight into molecular level protection mechanisms of molecular adhesion promotors during anodic polarization.

\section{Experimental Section}
\subsection*{Chemicals and Sample preparation}
Electrolyte solutions were prepared from $NaClO_4$ (98\%, Alfa Aesar) and Milli-Q water (resistivity \textgreater 18 M$\Omega\cdot$ cm, total organic carbon < 4 ppb).
11-Mercapto-1-undecanol (97\%, Sigma Aldrich),
1-Undecanthiol (98\%, Sigma Aldrich) were diluted to 1~mM solutions in Ethanol (chromatography grade, Carl Roth) for coating of the substrates.
Molecularly smooth gold surfaces of 70-100~nm thickness were prepared by template stripping from mica using an established protocol \cite{Hegner1993Jul, Chai.}.
Samples are contacted with 0.125~mm diameter gold wire (99.99\%, Goodfellow Cambridge Ltd.), and immersed for min. 12~h in ethanolic SAM solution. To remove excess thiols surfaces are rinsed thoroughly with ethanol, hexane and ethanol and dried with a gentle $N_2$ stream.
% 70-100~nm gold (99.99\%, MaTecK) were prepared with physical vapor deposition (home-built system) on mica  template-stripped on glass slides

\subsection*{Atomic Force Microscopy}
AFM topographies were taken with a Cypher ES (Asylum Research, Oxford Instruments, Santa Barbara, US) using Arrow\textsuperscript{TM} UHFAuD (NanoWorld, CH) and SCOUT 350 RAu (NuNano, GB) probes. Photothermal excitation is used with amplitude modulation (blueDrive) as driving mode. The electrochemical cell is a home-built modification with a platinum foil as a counter electrode and the reference electrode being connected via a capillary to the enclosed cell.

\subsection*{Inductively Coupled Plasma Mass Spectrometry}
Measurements were carried out with an Agilent 7900 ICP-MS (Agilent Technologies, US), a collision cell with 5~mL/min flow of helium as cell gas was used. Calibration was performed with multi-element standard (Inorganic Ventures, US). 
For ICP-MS, downstream of the electrochemical cell, the analyte was mixed with internal standard solution, containing Cobalt an Thallium. Electrochemical experiments were conducted in a home-built flow cell out of PEEK and PTFE \cite{Dworschak.2020}.
The exposed electrode area is circular and sealed with a 3 mm inner diameter O-ring. 
Before each experiment the electrolyte was purged with compressed and filtered air for at least 30 minutes to guarantee the same concentration of dissolved oxygen. 
\subsection*{Electrochemistry}
A Biologic VSP-300 potentiostat (Biologic, France) or a PalmSens 4 (PalmSens BV, NL) were used for electrochemical measurements.
All electrochemical experiments were performed with a Ag|AgCl-Electrode (Multichannel Systems, DE) as reference electrode. All presented data is referenced to that potential. Platinum is forming the counter electrode. Freshly prepared SAMs on gold get - after a short equilibrating time in the electrolyte (10~mM $NaClO_4$ for AFM, 100 mM for ICP-MS) - polarized from 0.0 to  1.5~V vs. Ag|AgCl with a scan rate of 5 mV/s. 

\section{Results and Discussion}
\begin{figure}
    \centering
    \includegraphics[scale=0.4]{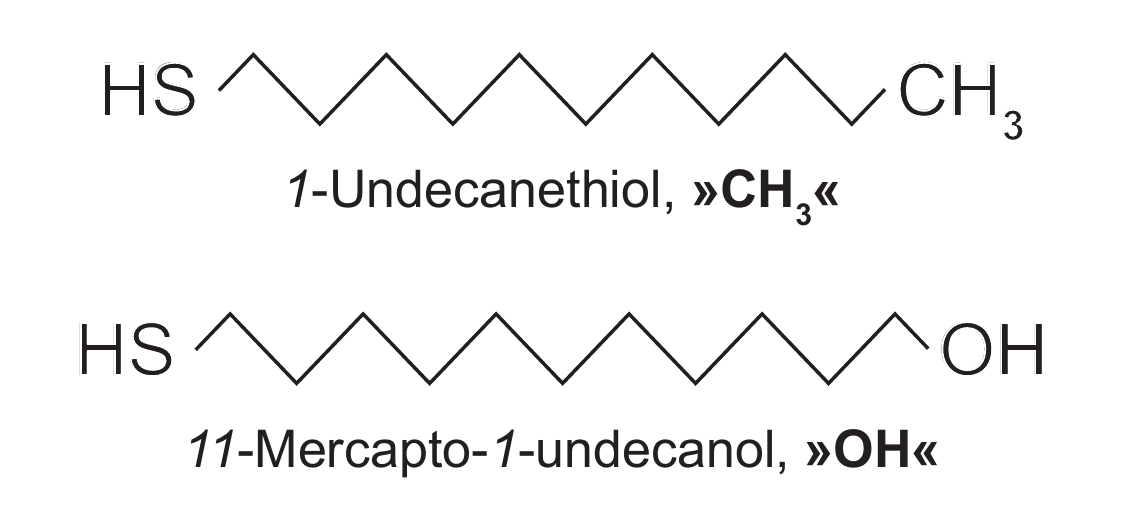}
    \caption{Thiols used for self-assembled monolayer films}
    \label{fig:f1}
\end{figure}

In this work we experimentally characterized the corrosion inhibition properties of hydrophilic and hydrophobic self-assembled monolayers (SAMs) on gold under potential control. As shown in \textbf{Fig.  \ref{fig:f1}}, we selected a hydrophobic and a hydrophilic head group termination, in order to study the effect of the wettability of the substrate. 

Therefore, the used SAMs are characterized by their same length of the hydrocarbon chain, but a major difference in interaction with water due to the selected head group. 
\textit{11-Mercapto-1-undecanol} ($OH$-SAM) with its hydroxyl-terminated tail is hydrophilic, whereas \textit{ 1-Undecanthiol} ($CH_3$-SAM) creates a hydrophobic layer on the gold. 
As \textbf{Fig. \ref{fig:f2} A$_1$} \& \textbf{B$_1$} show, the preparation produced - as expected from literature \cite{Shaheen.2017,Kakiuchi.2001,Moores.2011,Xu.1998} - an uniform and smooth film with typical defect patterns at domain boundaries of the formed SAMs. 

These SAMs were consequently anodically polarized to 1.5~V in order to characterize their corrosion inhibiting behaviour during anodic polarization using Linear Sweep Voltammetry (LSV). The data was further complemented by \textit{in situ} and \textit{ex situ} AFM topography scanning, as well as by online ICP-MS flow cell analysis of the anodically dissolving gold. The results of these analyses can be summarized as follows: 

\begin{figure}
    \centering
    \includegraphics[scale=0.7]{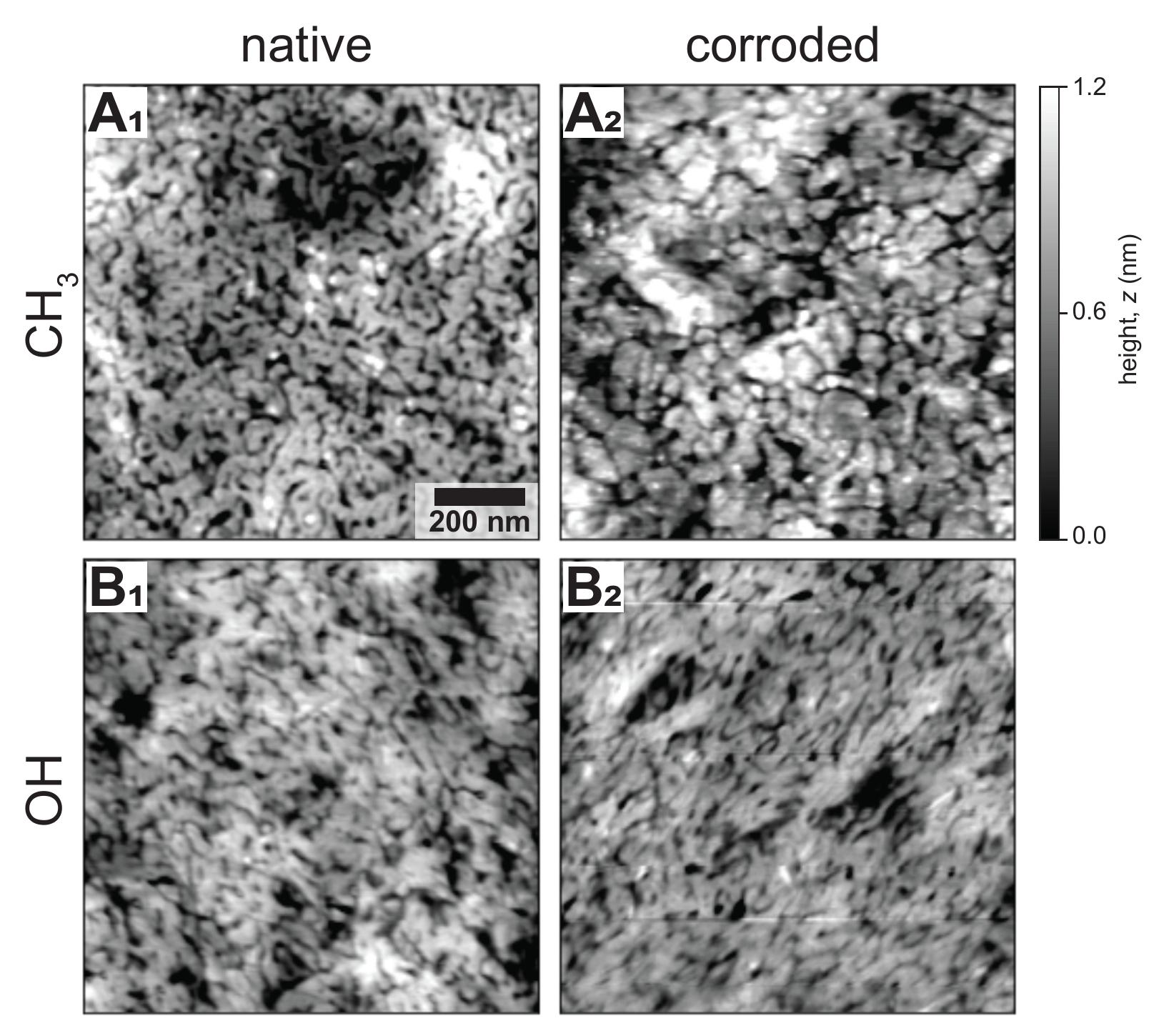} %width=\textwidth
    \caption{\textit{Ex situ} AFM topographies of self assembled monolayers on gold before (1) and after (2) polarisation to 1.5~V for $CH_3$- (A) and $OH$- (B) SAMs. }
    \label{fig:f2}
\end{figure}

First, \textit{ex situ} topographies (\textbf{Fig. \ref{fig:f2}}) compare the initial conditions of the surfaces modified with the respective SAMs (indicated by subscript $_1$) with a topography after polarization.
For the hydrophobic $CH_3$-SAM the surface after polarization (\textbf{$A_2$}) shows two topographic features, which are (1) globular residues likely formed by disintegrated SAM, as well as (2) nano-scaled grains as expected for a neat template stripped gold surface. 
This behaviour suggests that the $CH_3$-SAM disintegrates completely during anodic oxidation. 
In contrast, the hydrophilic SAM still shows similar structures before and after polarisation, indicating an almost unaltered and likely intact SAM. 

Hence, the head group directly determined how the SAM behaves during polarization. 
It appears likely, that the thiol|gold bond is weakened by the growth of an underlying oxide film, which in turn can cause a disintegration of the hydrophobic SAM. 
This disintegration can be driven by the head group effect as follows: Weakening of the surface bond can trigger the formation of bilayer structures, or can even drive micelle/particle formation as observed for weak binding of phosphonate SAMs on alumina \cite{Thissen.2010}, if the hydrophobic interaction of the head groups overpowers the thiol|gold binding. 
In contrast, if the SAM molecules are terminated by hydrophilic head groups there is no direct driving force for a disintegration of the SAM, and the intramolecular hydrophobic interactions within the SAM can withstand the "rolling up"/ lift-off of the SAM. As such, for the $OH$-SAM, the weakening of the thiol bond may result in a "flying carpet like" situation during anodic polarization.

\begin{figure}
    \begin{center}
    \includegraphics[scale =0.7]{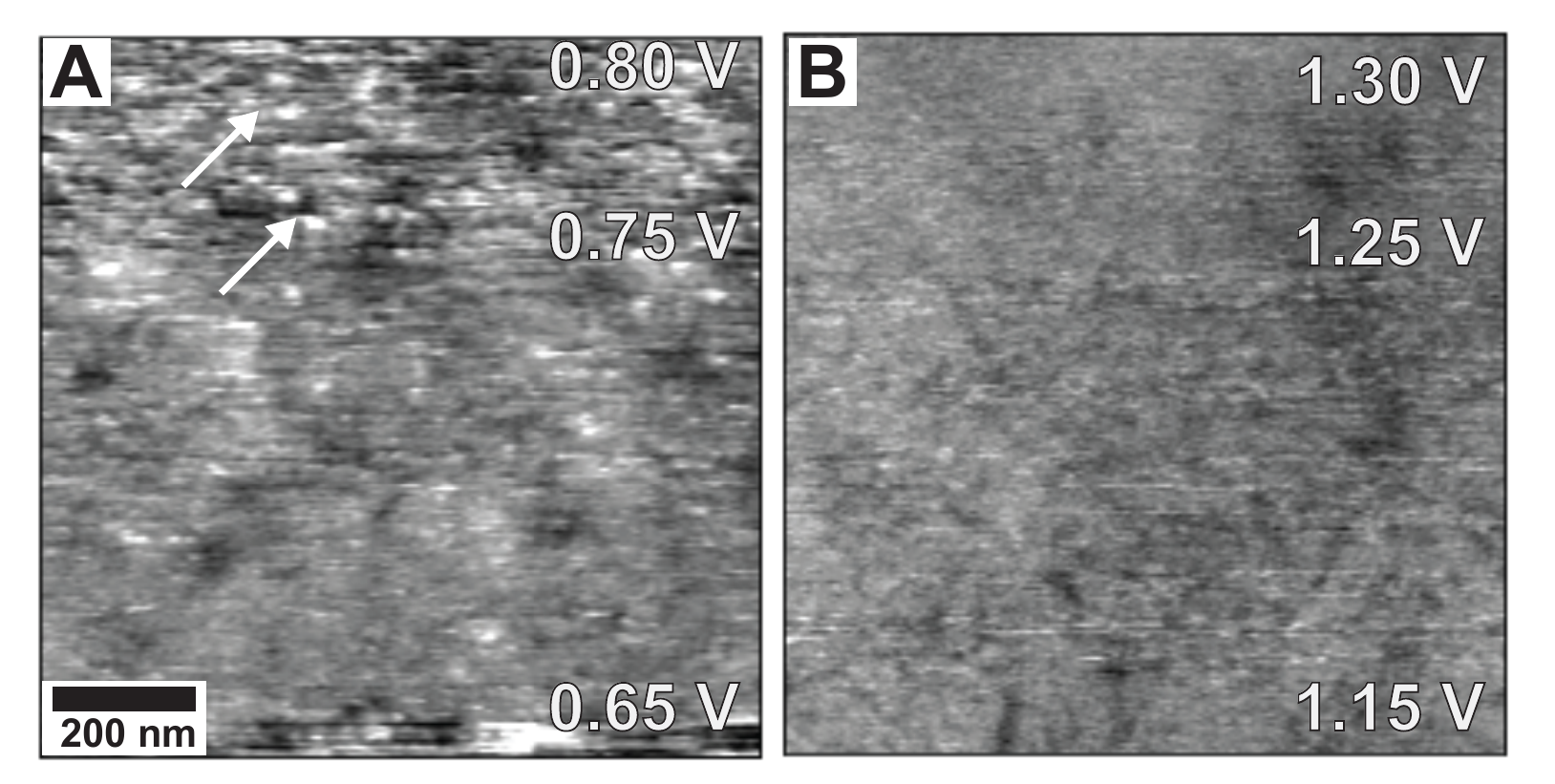}
    \caption{\textit{In situ} AFM topographies of A) $CH_3$- and B) $OH$-SAM on gold during linear polarization.}
    \end{center}
    \label{fig:f3}
\end{figure}
To further support this interpretation, we performed \textit{in situ} AFM under potential control. 
\textbf{Fig. \ref{fig:f3}} shows an image of the representative scan, when a significant change of the surface topography was observed during anodic polarization. 
Videos of the entire topographic evolution during polarization are available for download as supporting information (SI). 

In detail, the hydrophobic SAM (\textbf{Fig. \ref{fig:f3} A}) is showing a roughening of the surface starting at 0.75~V, in terms of formation of particulate residues (marked by arrows in figure). 
This is consistent with the interpretation of a weakening of the thiol|gold bond and a consequent triggering of a SAM lift-off due to formation of micellar structures, which is driven by the hydrophobicity of the molecules. The roughening of the $CH_3$-SAM hence indicates the initial detachment of the thiols accompanied by micelle formation. 
In contrast, the hydrophilic SAM appears intact up to more than 1.2~V where we see considerable flattening of the surface.
This flattening is indicative of a lift-off of the SAM as a "flying carpet", i.e. the thiol-gold bond is weakening, and as a results a more mobile, less defect rich SAM structure forms, and does also not impregnate the underlying granular structure of the gold. 
As seen in the \textit{ex situ} data (see again \textbf{Fig. \ref{fig:f2} B$_2$}) the SAM remains then intact and reforms after polarization, without lifting off from the surface. 

Based on this nanoscopic understanding of the SAM behaviour during polarisation we performed additional \textit{in situ} spectroelectrochemical analysis \cite{Cherevko.2013,Dworschak.2020,Ogle.2000}, to understand how these SAM structures inhibit or enhance gold dissolution during oxidative polarization. 

\begin{figure}
    \centering
    \includegraphics[scale=0.4]{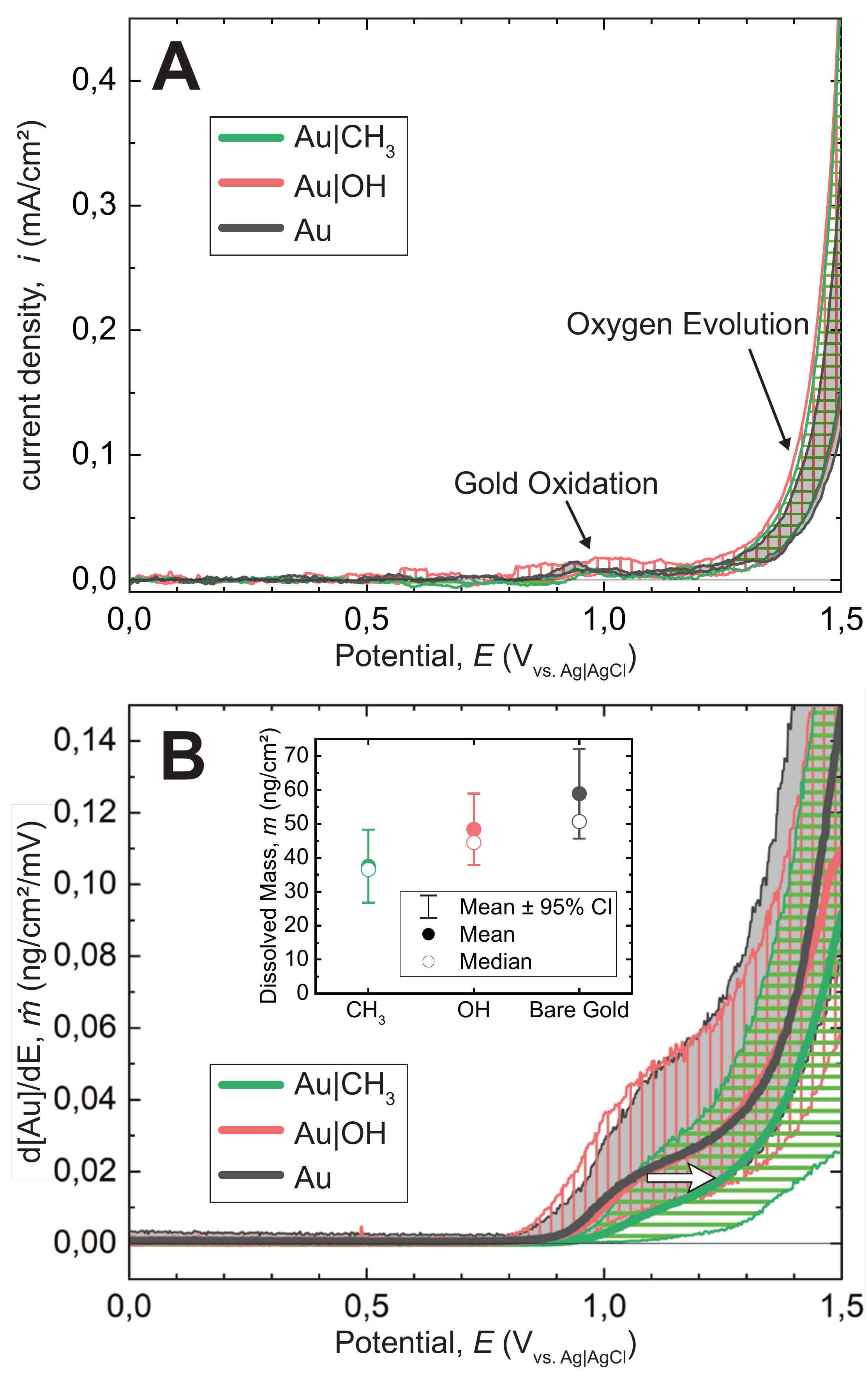}
    \caption{Linear Sweep Voltammetry (LSV) of bare gold (black), gold covered with a hydrophilic (red) and hydrophobic (green) self-assembled monolayer. Shaded areas indicate the range of observed dissolution profiles over at least nine experiments.
    \textbf{A}) Electrochemical Polarization Curves.
    \textbf{B}) Dissolved Gold during LSV.
    Solid lines represent the average curves over all of one type.
    Inset shows total integrated dissolved mass of gold during LSV, whiskers indicate the confidence interval (CI) of 95\%.}
    \label{fig:f4}
\end{figure}

\textbf{Fig. \ref{fig:f4}} shows (a)~the measured current during electrochemical polarization as well as (b)~the elementally resolved dissolution current of the gold dissolution (ICP-MS) displayed as a function of the linearly increasing potential.  
For both, data of more than 8 independent measurements are compiled for each SAM sample ($OH$ and $CH_3$) and are further compared to the data obtained for bare gold. For the elemental dissolution current (ICP-MS) shown in panel \textbf{B} the observed range (shaded areas) as well as the mean dissolution current are displayed (solid lines). For the current, two examples are shown, as less variation is observed.  

The electrochemical current shown in panel \textbf{A} and as marked indicates the inital gold oxidation at the expected potential, as well as the increasing current at higher polarizations due to water splitting. 
Statistically, there is no large difference in the absolute observed currents. 

However, the ICP-MS dissolution currents of gold exhibit clear changes, and an inhibition of the anodic dissolution of gold for the $CH_3$-SAM and the $OH$-SAM (shown in \textbf{Fig. \ref{fig:f3} B}) compared to unmodified gold is obvious. 
In detail, both SAMs statistically shift the onset of rapid gold dissolution to higher potentials. 
It is worth noting that the recorded dissolution rates show a significant variation although preparation remains the same, indicating an influence of defects formed during SAM formation. 
Nonetheless, statistical trends hold over a large set of >8 experiments for each sample. 

All of the curves have in common a two-step dissolution profile with a small shoulder of comparatively little dissolution starting with the gold oxidation, followed by a steep increase in amount of gold released to solution during water splitting. 
Previous work by Cherevko et al. \cite{Cherevko.2013} traced this behaviour back to initial small dissolution from formation of gold oxide and later on more drastic degradation during water splitting, which is consistent with our data of SAM coated substrates. 
%%% INSET with total amount of dissolved Gold

The inset in \textbf{Fig. \ref{fig:f3} B} shows the integrated amount of dissolved gold, further confirming this trend. 
%It has to be noted that integration of total dissolved gold also includes the gold still dissolved after the end of the polarization, since time-resolution of the used setup is influenced by diffusion of analyte on the way from electrode to detection. 
Dissolution for all of the systems tested correspond to only a fraction of a single layer of gold which would result in a total of $\approx$~450~ng/cm² gold dissolution. 
As expected the uncoated gold shows the highest dissolved mass, $OH$- as well as $CH_3$-SAM modified gold show a decreasing trend to about 30\% lower overall gold dissolution. 
The SAM coating can hence significantly suppress gold dissolution, likely by a barrier effects, and potentially also by stabilisation of the surface atoms at the SAM|oxide interface. 

Surprisingly, the least gold is released from the $CH_3$-coated surface, although the SAM lift-off is observed at the lowest potentials in AFM and the SAM indicates a significant roughening.
Further, looking at the ranges of dissolution recorded (indicated by the shaded areas)  the hydrophobic SAM (green) clearly shows an onset of gold dissolution at higher potentials compared to hydrophilized and bare gold.
Bare gold and $OH$ coated gold overlap over most of the potentials, just for very high potentials at above 1.4~V gold with $OH$-SAM seems to be slightly less dissolving.
The average dissolution curve for the $CH_3$-SAM rises later and for the whole potential range stays well below the dissolution rates of the other systems tested.
As such, and in agreement with our interpretation, the hydrophobic $CH_3$-SAM may form micelles, which may lead to a trapping of gold within micelles, hence lowering the total dissolved ion count initially.
This initial low count is followed by the steepest rise of the dissolution, at higher potentials, where trapped micelles may desorb. 

This is an interesting behaviour, which we can interpret as a potentially important fundamental step during technical processes such as interphase formation during polymer coating (gluing/ coating for corrosion protection, etc.) of a metal in an oxidative environment.  Interphases are considered boundary layers of a metal|polymer interphase where it has been speculated that metal ions dissolve and stabilize into the polymer matrix due to their interaction with the functional groups of the polymer. 
Our data demonstrates that such a metal dissolution mechanism is possible, and appears to be favoured by intermolecular interactions that drive micelle formation, or in other words the enclosure of metal ions in an functional organic matrix that is in contact with the metal during oxidation. 

\section{Conclusions}
We successfully compared AFM imaging and ICP-MS flow cell studies of the anodic detachment of a protective thin film on gold. 
We demonstrated their corrosion-inhibiting effect at anodic potentials. 
The morphological changes of two SAMs (11-Mercapto-1-undecanol and 11-Undecanethiol) on gold during anodic polarization show a strong effect of the head group in \textit{in situ} AFM. 
Quantification of the dissolved gold with a scanning-flow cell coupled to an ICP-MS certainly adds information gain to the pure electrochemical data and might prove useful for the study of further coatings on metals.  

Our results suggest the following specific conclusions:
\begin{itemize}
    \item Hydrophobic molecules may immobilize dissolving metal ions in micelles, which may enclose metallic (or oxidic) nanoparticles formed during anodic oxidation.
    \item During this process the SAM disintegrates and dewetting from the oxidized gold interface occurs. 
    \item This data may show initial fundamental steps which occur during interphase formation when a coating or glue is applied to a metallic substrate.
    \item Hydrophilic molecules tend to not detach, but form a weakly adhering and highly mobile layer without significant micelle formation. This results in lower initial retardation of metal dissolution. 
\end{itemize}

This combinatorial multi-technique approach will prove useful for studying the interfacial activity and corrosion suppression mechanism of inhibiting molecules on other metals and alloys.

\section*{Acknowledgements}
The authors acknowledge support by the European Research Council (Grant: CSI.interface, ERC-StG 677663, characterization of surfaces). M.V. and C.B. gratefully acknowledge support from the K1-COMET center CEST (Centre for Electrochemical Surface Technology, Wiener Neustadt) funded by the Austrian Research Promotion Agency (FFG).

\section*{Competing Interests}
The Authors declare no Competing Financial or Non-Financial Interests. 

\section*{Data availability}
The raw and processed data required to reproduce these findings are available from the corresponding author via www.repositum.tuwien.ac.at upon reasonable request.
\section*{Author Contributions}
\textbf{Dominik Dworschak:} Investigation, Methodology, Visualization, Writing-Original Draft.
\textbf{Carina Brunnhofer:} Investigation, Data Curation.
\textbf{Markus Valtiner:} Conceptualisation, Writing-Reviewing and Editing, Funding acquisition, Supervision.

% \typeout{}
\bibliographystyle{elsarticle-num}  
\bibliography{references}  
%%% Remove comment to use the external .bib file (using bibtex).

\end{document}